# Economic and Technological Complexity:
## A Model Study of Indicators of Knowledge-based Innovation Systems

Inga Ivanova,* [a] Øivind Strand,[b] Duncan Kushnir,[c] and Loet Leydesdorff [d]

**Abstract**

The Economic Complexity Index (ECI; Hidalgo & Hausmann, 2009) measures the complexity of national economies in terms of product groups. Analogously to ECI, a Patent Complexity Index (PatCI) can be developed on the basis of a matrix of nations versus patent classes. Using linear algebra, the three dimensions—countries, product groups, and patent classes—can be combined into a measure of "Triple Helix" complexity (THCI) including the trilateral interaction terms between knowledge production, wealth generation, and (national) control. THCI can be expected to capture the extent of systems integration between the global dynamics of markets (ECI) and technologies (PatCI) in each national system of innovation. We measure ECI, PatCI, and THCI during the period 2000-2014 for the 34 OECD member states, the BRICS countries, and a group of emerging and affiliated economies (Argentina, Hong Kong, Indonesia, Malaysia, Romania, and Singapore). The three complexity indicators are correlated between themselves; but the correlations with GDP per capita are virtually absent. Of the world's major economies, Japan scores highest on all three indicators, while China has been increasingly successful in combining economic and technological complexity. We could not reproduce the correlation between ECI and average income that has been central to the argument about the fruitfulness of the economic complexity approach.

---

[a] * corresponding author; Institute for Statistical Studies and Economics of Knowledge, National Research University Higher School of Economics (NRU HSE), 20 Myasnitskaya St., Moscow, 101000, the Russian Federation; inga.iva@mail.ru
[b] Norges teknisk-naturvitenskapelige universitet (NTNU), Department of International Business, Larsgårdsvegen 2, 6009 ÅLESUND, Norway; oivind.strand@ntnu.no
[c] Chalmers University of Technology, Göteborg 412 58, Sweden; kushnir@chalmers.se
[d] Amsterdam School of Communication Research (ASCoR), University of Amsterdam, PO Box 15793, 1001 NG Amsterdam, the Netherlands; loet@leydesdorff.net

## 1. Introduction

Hidalgo & Hausmann (2009) proposed the Economic Complexity Index (ECI) using the portfolios of countries in terms of product groups which they export to quantify a country's economic complexity. A country's economic growth and income can be expected to depend on the diversity of the products in its portfolio (Cadot *et al*., 2013). Given the two axes of the matrix of countries versus product groups, Hausmann *et al.* (2011, p. 24) also introduced the product complexity index (PCI) which measures the spread of the production of each product group over nations. The complexity of a country's economy, in turn, refers to the set of capabilities, accumulated by that country.

According to Hidalgo & Hausmann (2009; henceforth HH) ECI is correlated with a country's income as measured by GDP per capita (Hidalgo & Hausmann, 2009: Fig. 3 at p. 10573). HH submit that the deviation of ECI from a country's income can be used to predict long-term future growth because a country's income can be expected to approach a competitive level associated with its economic complexity (Ourens, 2013, p. 24).[5] Hence, ECI could be considered as a predictive measure of a country's competitive advantage in the future.

Since based on the product portfolios, ECI values can be expected to reflect the manufacturing capabilities of countries (Hausmann *et al*, 2011, p. 7). However, HH did not provide an explicit definition of the manufacturing capabilities and their respective knowledge bases. In our opinion, manufacturing complexity is inevitably related to the knowledge intensity and sophistication of exports of products with comparative advantages (e.g., Foray, 2004; Foray & Lundvall, 1996; OECD, 1996; ECR, 2013). One needs an advanced indicator of competitiveness which indicates whether manufacturing industries in a country have a relatively high degree of complexity.

---

[5] Kemp-Benedict (2014) noted that the correlation between income and ECI can also be considered as a consequence of the well-known relation between export and income growth.

New industries are more likely to be generated in regions where they can be technologically related to existing industries (Boschma *et al*., 2013; Frenken *et al*., 2007; Neffke *et al*., 2011). Although regional diversification is often studied in terms of industrial dynamics, specification of the technological (knowledge) dynamics would enable us to make a direct link between urban diversification and technology portfolios. Boschma *et al*. (2014, at p. 225), for example, concluded from a study of 366 US cities during the period 1981-2010 that "technological relatedness at the city level was a crucial driving force behind technological change in US cities over the past 30 years."

Arguing that the knowledge dimension is "intangible," Cristelli *et al* (2013) proposed to model capabilities as a hidden layer between products and countries. In a series of studies, Luciano Pietronero and his colleagues (e.g. Cristelli *et al*., 2015; Tacchella *et al*., 2013) have further developed this alternative model of economic complexity from a data-driven perspective. The resulting models predict GDP and other economic parameters in much detail. From the perspective of innovation studies, however, there remains a need for an *explicit* measure of the technological capabilities of nations. Can the missing link between product groups and technology (patent) portfolios be endogenized into the model (Nelson & Winter, 1977, 1982) instead of being handled as a residual (Solow, 1957) or latent factor? Proponents of endogenous growth theory, for example, have argued that economic growth is the result of combinations of technologies and manufacturing (Romer, 1986). The longer-term research question is how to compare (national) systems of innovation in terms of their efficiency in coupling the global dynamics of markets and technologies at the level of firms, institutions, and nations (Freeman & Perez, 1988; Lundval, 1988 and 1992; Nelson, 1992; Reikard, 2005).

In this study, we address this question step by step. In addition and analogously to HH's product diversity, the technological diversity of a country can be measured, for example, in terms of patent portfolios. Patents have been considered as a measure of innovative activity in the innovation studies literature (e.g., Arcs & Audretsch, 2002), although patents are indicators of invention, not innovation. However, it is less problematic to consider patents as indicators of the dynamics of technological knowledge (Alkemade *et al*., 2015; Verspagen, 2007). Patents can also be strategic (Blind *et al*., 2006; Hall & Ziedonis, 2001; cf. Jaffe & Trajtenberg, 2002).

Using the patent portfolio as a proxy for the technological complexity of a country, we first develop the Patent Complexity Index (PatCI; cf. Balland *et al*., 2016). We then use patent-product concordance tables to construct a third matrix of product groups versus technology classes. In a three-partite network of relations among countries, product groups, and patent categories, each third category can be expected to provide feedbacks or feed-forwards on the relation between the other two. The feedbacks and feed-forwards generate loops that can provide new options, synergies, and integration (Petersen *et al*., 2016). The endogenization of the technological dimension in a three-partite network will enable us to derive a "Triple Helix"-type indicator for the measurement of relative integration in national systems of innovation.

Since the model is developed at the macro-level of nations, the empirical elaboration can be policy relevant at that level. We follow HH's choice for data at this macro-level. Our model is therefore not micro-founded. From a formal perspective, however, one can similarly (alternatively) study the relations among firms, product groups, and patent classes as another empirical domain; but using the same algorithms. More generally, one can argue that positive feedback in the cycling among three dimensions models the potential synergy in the interactions, whereas negative feedback models a form of institutional lock-in. In empirical cases, both processes can be expected to operate simultaneously. Accordingly, the Triple Helix Complexity Indicator (THCI) derived below evaluates the resulting configuration by aggregating two dynamics: the organizational and integrating dynamics of localized retention and the self-organizing dynamics of markets and the techno-sciences as globalizing selection environments (Leydesdorff *et al*., 2017, in press). One can also consider the cycling as a form of auto-catalysis that has the potential to bi-furcate and thus develop long-term cycles (Ulanowicz, 2009; cf. Ivanova & Leydesdorff, 2015).

In summary, this study aims to extend ECI in the technological dimension and then integrate the model across the three dimensions. Our first contribution is to derive the other two indicators (PatCI and THCI) and their relationships to ECI. Secondly, the empirical results raise questions for future research. For example, HH's choice for the Revealed Comparative Advantage index (RCA; Balassa, 1965) may be unfortunate from the perspective of complexity analysis and indicator development. Whereas RCA is firmly embedded in classical (Ricardian) trade theory, one binarizes the matrix and thus throws away valuable information about a country's comparative advantages in products or technologies. A valued measure may much improve the indicator when compared with a binary one.

The paper is structured as follows: Section 2 first provides the derivation of ECI. We then specify the analogous construction of the Patent Complexity Index (PatCI), generalize HH's so-called Method of Reflections (MR) to three (or more) dimensions, and derive the Triple Helix

Complexity Index (THCI). Section 3 describes the data collection and Section 4 presents the empirical results. The main findings and conclusions are summarized in Section 5.

## 2. Methods

### a) Economic Complexity Index

HH's ECI is derived from a matrix $M_{c,p}$ where the index $c$ refers to a country and $p$ refers to a product group. The matrix elements are assumed to be one if Balassa's (1965) RCA is larger than or equal to one and otherwise zero:

$$RCA_{c,p} = \frac{x_{c,p} \big/ \sum_p x_{c,p}}{\sum_c x_{c,p} \big/ \sum_{c,p} x_{c,p}} \qquad (1)$$

where $x_{c,p}$ is the value of product $p$ manufactured by country $c$. According to HH (at p. 10571) "a country can be considered to be a significant exporter of product p if its Revealed Comparative Advantage (the share of product $p$ in the export basket of product $p$ in world trade) is greater than 1" (Hidalgo and Hausmann, 2009, p. 10571).

Summing the elements of matrix $M_{c,p}$ by rows (countries), one obtains a vector with components referring to the corresponding products and indicating a measure of product *ubiquity* relative to the world market. The sum of matrix elements over the columns (products) provides another vector defining the *diversity* of a country's exports:

$$k_{p,0} = \sum_{c=1}^{N_c} M_{c,p}$$
$$k_{c,0} = \sum_{p=1}^{N_p} M_{c,p} \qquad (2)$$

Where $N_c$ is defined as the number of countries and $N_p$ as the number of product groups—HH use $N_c =178$ and $N_p=4948$; see section 3 below—more accurate measures of diversity and ubiquity can be obtained by adding the following iterations:

$$k_{p,n} = \frac{1}{k_{p,0}} \sum_{c=1}^{N_c} M_{c,p} k_{c,n-1}$$
$$k_{c,n} = \frac{1}{k_{c,0}} \sum_{p=1}^{N_p} M_{c,p} k_{p,n-1}$$

(3)

HH (at p. 10571) call this "the method of reflections" (MR): each product is weighted proportionally to its ubiquity on the market, and each country is weighted proportionally to the country's diversity. Substituting the first equation of system (3) into the second, one obtains:

$$k_{c,n} = \frac{1}{k_{c,0}} \sum_{c'=1}^{N_c} \sum_{p=1}^{N_p} M_{c,p} \frac{1}{k_{p,0}} M_{c',p} k_{c',n-2}$$

(4)

Because empirically the sequence $k_{c,n}$ converges to a limit equation (4) can be formulated as a matrix equation:

$$\vec{k} = W \cdot \vec{k}$$

(5)

where vector $\vec{k}$ is a limit of iterations, as follows:

$$\vec{k} = \lim_{n \to \infty} k_{c,n}$$

(6)

HH introduce the economic complexity index (ECI) as an eigenvector $\vec{k}$ of the matrix $W_{c,c'}$

$$W_{c,c'} = \sum_p \frac{M_{c,p} M_{c',p}}{k_{c,0} k_{p,0}}$$

(7)

associated with the second largest eigenvalue, because it can be shown mathematically that in this case eigenvectors associated with the second largest eigenvalue capture most of the variation (Kemp Benedict, 2014). ECI is then defined according to the formula

$$ECI = \frac{\vec{k} - <\vec{k}>}{stdev(\vec{k})} \qquad (8)$$

ECI is a vector of which the components refer to the respective countries.

### b) Patent Complexity Index

HH hypothesize that diversity and ubiquity scores of the countries reflect underlying "capabilities." By capabilities they imply the ability of countries to make corresponding products; but this concept can also be extended to technologies. The corresponding technologies are legally documented as patents. Patents can be used as a proxy measure for technological capabilities. Using an analogous design, one can construct a matrix $M_{c,t}$, which is essentially matrix $M_{c,p}$ in which the product groups, indicated by index $p$, are substituted by patent technology classes, indicated by index $t$ (Balland *et al.*, 2016). Following the MR formalism explained above (Eqs. 2-8), one can derive a matrix $M_{c,c'}$, equivalent to $W_{c,c'}$ in Eq. 7, as follows:

$$M_{c,c'} = \sum_t \frac{M_{ct} M_{c't}}{\rho_{c,0} \rho_{t,0}} \qquad (9)$$

and the Patent Complexity Index (PatCI) is estimated in accordance with Eq. (8), as follows:

$$PatCI = \frac{\vec{k} - <\vec{k}>}{stdev(\vec{k})} \qquad (10)$$

The condition for the RCA index is in this case:

$$RCA_{c,t} = \frac{x_{c,t} / \sum_c x_{c,t}}{\sum_t x_{c,t} / \sum_{c,t} x_{c,t}} \qquad (1')$$

where $x_{c,t}$ is a number of patents possessed by a country $c$ in a patent group $p$ counted on the integer or fractional base. This condition is met when the weight of a technology group in a country's portfolio exceeds the average weight in the set. In other words, the country is specialized in this specific technology. Diversity and ubiquity scores (Eq.3) would reflect each country's technological diversification and the prevalence of a particular technology in its portfolio, respectively.

In summary, PatCI captures the technological diversification of a country expressed in terms of patent portfolios. Note that this measure is volatile when applied to small and less developed countries as compared with more developed ones, because in the case of small and less developed countries small changes in the number of patents may lead to disproportionate changes in PatCI. For this reason, we will limit the presentation of the measurement results of PatCI (in Section 4) to large and medium-sized economies.

*c) The three complexity sets*

HH (2009, p. 10570) noted that "the bipartite network connecting countries to products is a result of tripartite network connecting countries to their available capabilities and products to the capabilities they require." However, ECI is a two-dimensional indicator, since the third (i.e., technological) dimension is not explicitly accounted for. After adding the technological dimension in terms of patent classes, this dimension can also be explicitly combined with the first two ones—countries and product groups. In addition to the matrices $M_{c,p}$ and $M_{c,t}$, one thus obtains a third matrix $M_{p,t}$ , in which index $p$ refers to product groups and $t$ to patent classes. In other words: a specific technology can be used in different product groups, and product groups can combine different technology classes. We thus infer that technologies can be related to products to variable extents.

As a matrix $M_{p,t}$ can be taken patent-product concordance table in which product groups are linked to patent classes (e.g., van Looy *et al*., 2015). The corresponding elements of the matrix $M_{p,t}$ are equal to 1 if product group $p$ comprises technology class $t$, and 0 otherwise. Note that this matrix may empirically be sparse, since many products are not related to patents and the distribution of patents over classes is very skewed.

Analogously to Eq. 2, one can define a product-technology diversity vector $\eta_{p,0}$ and a technology-ubiquity vector $\eta_{t,0}$ as follows:

$$\eta_{p,0} = \sum_t M_{p,t}$$
$$\eta_{t,0} = \sum_p M_{p,t}$$

(11)

In Eq. 11 $\eta_{p,0}$ represents technological sophistication of a product (i.e., how many different technologies are comprised in the product). Combining more technologies in a product makes it more "complex": $\eta_{t,0}$ measures the ubiquity of a technology over different product groups.

Following the iterative procedure described above, one is now able to construct three groups of vectors:

$$k_{p,0} = \sum_{c=1}^{N_c} M_{c,p}$$
$$k_{c,0} = \sum_{p=1}^{N_p} M_{c,p}$$

(12)

$$\rho_{c,0} = \sum_{t=1}^{N_t} M_{c,t}$$
$$\rho_{t,0} = \sum_{c=1}^{N_c} M_{c,t}$$

(13)

$$\eta_{p,0} = \sum_{t=1}^{N_t} M_{p,t}$$
$$\eta_{t,0} = \sum_{p=1}^{N_p} M_{p,t}$$

(14)

which are connected by the following iterative sequences:

1)     in the country-product dimension:

$$k_{p,n} = \frac{1}{k_{p,0}} \sum_{c=1}^{N_c} M_{c,p} k_{c,n-1}$$
$$k_{c,n} = \frac{1}{k_{c,0}} \sum_{p=1}^{N_p} M_{c,p} k_{p,n-1}$$

(15)

2)     in country-technology dimension:

$$\rho_{c,n} = \frac{1}{\rho_{c,0}} \sum_{t=1}^{N_t} M_{c,t} \rho_{t,n-1}$$
$$\rho_{t,n} = \frac{1}{\rho_{t,0}} \sum_{c=1}^{N_c} M_{c,t} \rho_{c,n-1}$$

(16)

3)     and in product-technology dimension:

$$\eta_{p,n} = \frac{1}{\eta_{p,0}} \sum_{t=1}^{N_t} M_{p,t} \eta_{t,n-1}$$
$$\eta_{t,n} = \frac{1}{\eta_{t,0}} \sum_{p=1}^{N_p} M_{p,t} \eta_{p,n-1}$$

(17)

where $k_{c,0}$ measures country product diversity, $k_{p,0}$ is product ubiquity over the set of countries, $\rho_{c,0}$ is country technological diversity, $\rho_{t,0}$ is technological ubiquity, $\eta_{t,0}$ is technological ubiquity with respect to products (i.e. how a specific technology is distributed across manufactured products), and $\eta_{p,0}$ is product technological sophistication (how many technologies are used in each product group). Each of these vectors can be associated with a corresponding complexity index. One thus obtains:

1. economic (ECI) and product group (PCI) complexity indices for the first group;

2. patent (PatCI) and technology (TCI) complexity indices for the second group;

3. product-technology (PTCI) and technology-product (TPCI) complexity indices for the third group.

The iterative couplings within each set generate three double-stranded helices corresponding to three bi-partite networks; but the interaction terms among the three helices are not yet included since these networks are not explicitly interconnected.

*d) Triple-Helix Complexity Index*

In the case of bi-lateral networks, described by Eqs. 15-17, the vector pairs are reciprocally interdependent, but the three pairs are not interacting. This is schematically depicted in Fig. 1a: the three groups of vectors are not connected to one another. In case of three-lateral network, however, we connect countries, technologies, and products in a cyclic manner, as depicted in Figs. 1b and c.

HH formulated as follows: "a country makes a product if it has all the necessary capabilities" and "the bipartite network connecting countries to products is a result of the tripartite network connecting countries to their available capabilities and products to the capabilities they require" (Hidalgo and Hausmann, 2009, p.10571). In other words, if a country possess or has access to a technology then it can produce the product, and vice versa - producing a product enables a country to further develop the corresponding technologies. In other words, these networks can be mapped as two cycles—country-technology-product and country-product-technology—which add to each other. If the clockwise cycle in Fig. 1b refers to a country-technology-product network, the counter-clockwise cycle in Fig. 1c refers to a country-product-technology network.

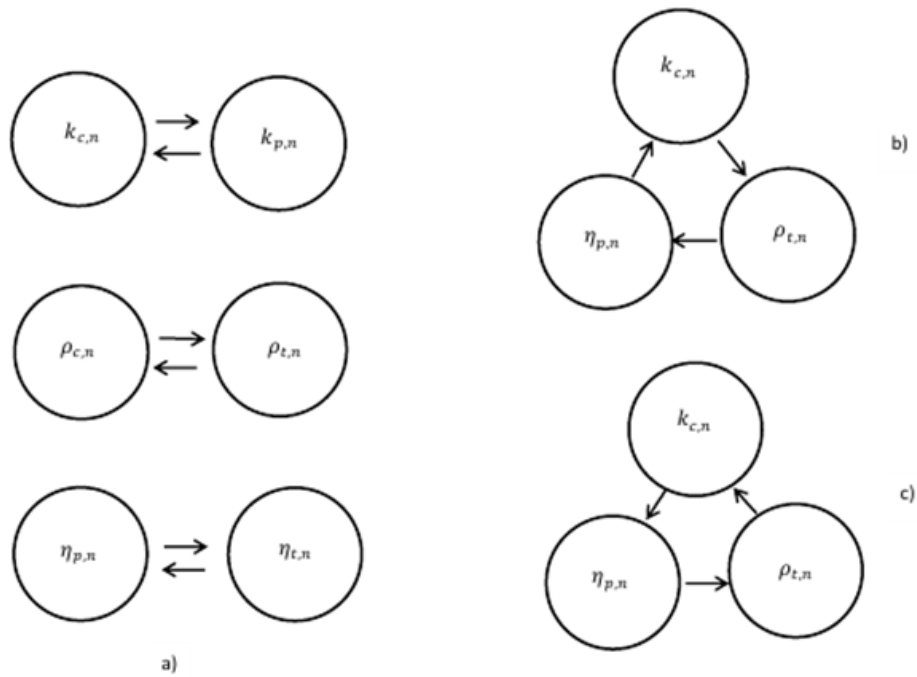

**Figure 1a, b, and c:** Reciprocal (a), cyclical clockwise (b) and counter-clockwise (c) interdependencies between complexity coefficients in iterative sequences.

The two cyclical configurations can analytically be distinguished, but operate empirically as feedback mechanisms on each other: local retention feeds back on the feed-forward in loops. This can be modeled as follows: three groups of vectors were distinguished in Eqs. 12-14 relating to the geographical ($k_{c,n}, \rho_{c,n}$), product ($k_{p,n}, \eta_{p,n}$), and technological ($\rho_{t,n}, \eta_{t,n}$) dimensions, respectively. One can extend this model with two groups of three vectors each relating to the other dimensions as follows:

1. clockwise (country-technology-product-country as in Fig. 1b):

$$k_{c,n} = \frac{1}{\rho_{c,0}} \sum_{t=1}^{N_t} M_{c,t} \rho_{t,n-1}$$

$$\rho_{t,n-1} = \frac{1}{\eta_{t,0}} \sum_{p=1}^{N_p} M_{p,t} \eta_{p,n-2} \qquad (18)$$

$$\eta_{p,n-2} = \frac{1}{k_{p,0}} \sum_{c'=1}^{N_c} M_{c',p} k_{c',n-3}$$

2. and counter-clockwise (country-product-technology-country as in Fig. 1c):

$$k_{c,n} = \frac{1}{k_{c,0}} \sum_{p=1}^{N_p} M_{c,p} \eta_{p,n-1}$$

$$\eta_{p,n-1} = \frac{1}{\eta_{p,0}} \sum_{t=1}^{N_t} M_{p,t} \rho_{t,n-2} \qquad (19)$$

$$\rho_{t,n-2} = \frac{1}{\rho_{t,0}} \sum_{c'=1}^{N_c} M_{c',t} k_{c',n-3}$$

The set of Equations (18) refers to clockwise cyclical interdependence (Fig. 1b); and the set of Equations (19) corresponds to counter-clockwise cyclical interdependence (Fig. 1c). In the case of cyclical interdependencies, each iterative step between two indicators is conditioned by the value of the third one. The two cycles (clockwise and counter-clockwise) operate in parallel, since they are coupled in the parameters.[6]

As noted, cyclical interdependence can also be considered as an auto-catalytic process (Ulanowicz, 2009); the alternative rotation provides the stabilizing feedback term to the globalizing feed forward of the auto-catalysis. Note that the cycles represent second-order relations among first-order relations, and not relations among the agents (nations, product groups, and patent classes) that are bilaterally related.

This structure of three mutually coupled dimensions shows an analogy with the Triple Helix (TH) model of innovations. In the TH model, three institutional actors (university, industry, and government) are expected to perform three functions: knowledge (technology) generation, (product) manufacturing, and legislative regulation respectively. Using a generalized TH model (e.g., Ivanova & Leydesdorff, 2014), the geographical dimension (countries) can be considered as a proxy for administrative regulation and legislation by government, technology classes as a proxy for the innovative knowledge dimension, and product groups as indicators of economic activity (cf. Petersen *et al*., 2016). In the TH model, the one cycle is associated with institutional organization and integration, and the other with "self-organization" and differentiation at the global level of markets and technologies (Leydesdorff & Zawdie, 2010). The trade-off between these two dynamics shapes a specific (e.g., national) system of innovation in terms of the efficiency of its integration and synergy (Petersen *et al*., 2006).

---

[6] Note that the country-related indicator $k_{c,n}$ is defined differently in Eqs. 20 and 21. In Eq. 20, a country's product diversity (i.e., how many different products the country manufactures) depends on technological ubiquity (how the technologies are distributed across the set of countries in the group) and technological ubiquity is conditioned by product technological diversity (how sophisticated are the products with respect to technologies which are used for their manufacturing). In the second case, the sequence is reversed and $k_{c,n}$ depends on product diversity and this relation is conditioned by technological diversity.

Each of the vectors in the iterative sequences defined by Eqs. (18) and (19) is modulated by the other vectors in cycles. After substituting the second and third equation from the system of Eq. (18) into the first one and eliminating $\rho_{t,n-1}$ and $\eta_{p,n-2}$ one obtains:

$$k_{c,n} = \frac{1}{\rho_{c,0}} \sum_{t=1}^{N_t} M_{c,t} \frac{1}{\eta_{t,0}} \sum_{p=1}^{N_p} M_{p,t} \frac{1}{k_{p,0}} \sum_{c'=1}^{N_c} M_{c',p} k_{c',n-3} \tag{20}$$

which can conveniently be written as a matrix equation:

$$\vec{k} = \boldsymbol{W} \cdot \vec{k} \tag{21}$$

where vector $\vec{k}$ is the limit value of iterations for $n \to \infty$.

$$\vec{k} = \lim_{n \to \infty} k_{c,n} \tag{22}$$

and matrix $\boldsymbol{W}$ has elements

$$W_{cc'} = \sum_{t=1}^{N_t} \sum_{p=1}^{N_p} \frac{M_{c,t} \cdot M_{p,t} \cdot M_{c',p}}{\rho_{c,0} \cdot \eta_{t,0} \cdot k_{p,0}} \tag{23}$$

In a similar way from Eq. 19 one can get:

$$\vec{k} = \boldsymbol{V} \cdot \vec{k} \tag{24}$$

where:

$$V_{cc'} = \sum_{t=1}^{N_t} \sum_{p=1}^{N_p} \frac{M_{c,p} \cdot M_{p,t} \cdot M_{c',t}}{k_{c,0} \cdot \eta_{p,0} \cdot \rho_{t,0}} \tag{25}$$

Proceeding in this way, the task of finding complexity coefficients in analogy to HH's argument, can be reformulated as a problem of linear algebra, and one can show that the maximum variability is captured by the eigenvector of $W$ with the largest eigenvalue less than one (Kemp-Benedict, 2014).

In other words, these three-mode networks can be considered as an elaboration of the TH model—operationalized in terms of countries (geography), product groups (industry), and patent classes (technological knowledge). The interactions among the three helices in the two cycles can be formalized as the Triple-Helix Complexity Index (THCI), as follows:

$$THCI = \frac{\vec{k} - <\vec{k}>}{stdev(\vec{k})} \qquad (26)$$

where $\vec{k}$ is a complexity vector obtained via summing complexity vectors obtained for the clockwise direction $\vec{k}(+)$ and the complexity vector obtained for the counter-clockwise rotation $\vec{k}(-)$, so that the evolution of the system can be defined as the result of interactions between the clockwise and counter-clockwise rotations, as follows:

$$\vec{k} = \vec{k}(+) + \vec{k}(-) \qquad (27)$$

THCI plays a unifying role in steering the complexity of products and technologies by adding complexity in the institutional (e.g., national) coupling (Freeman & Perez, 1988). Since the couplings evolve in terms of second-order relations, next-order cycles (e.g., technological regimes) can be expected to operate on the observable relations (e.g., technological trajectories; Dosi, 1982).

### 3. Data

We use the same data source for the measurement of ECI as HH, namely international trade data among nations made available by the UN Comtrade database at http://comtrade.un.org/data. Because our objective is not to further develop or refine ECI, we limit the set pragmatically to 45 relatively developed countries, including the 34 OECD member states,[7] the five BRICS countries, and Argentina, Hong Kong, Indonesia, Malaysia, Romania, and Singapore as emerging economies.[8] We collected data for these 45 countries for the period 2000-2015, according to the Standard International Trade Classification (SITC) Revision 3 at the three digit level; this matches export reports for 260 products. HH used 4,948 products at the four-digit level of Rev. 4 for the period 1992-2000, and all 178 countries. Because of the different delineations, the variation is different, but the design is similar. Our results can be considered as a partial, yet updated replication of the *Atlas of Economic Complexity* (Hausmann *et al*., 2011).

Data on patents for the same set of 45 countries and for the same period (2000-2014) were downloaded from the website of the U.S. Patent and Trade Organization (USPTO) as bulk data cache maintained by Google (at https://www.google.com/googlebooks/uspto.html). The International Patent Classifications (IPC) provide a fine-grained index system of patents worldwide that has been further developed in collaboration with the USPTO and the European Patent Organization (EPO) into the system of Cooperative Patent Classifications (CPC).[9] The system is elaborated to the level of 14 digits, although in our study we use the 129 classes at the 3-digit level as indicators of the technological dimension (however imperfect).[10] We use USPTO because patents at USPTO have been considered as more competitive for emerging markets than patents filed with other national or regional patent offices (Criscuolo, 2004; Jaffe & Trajtenburg, 2002). Patents are assigned fractionally to countries according to inventor addresses.

---

[7] Taiwan is not included because it is not a member-state of the United Nations.
[8] Argentina, China, Romania, Russia, Singapore, South Africa, and Taiwan are affiliated member economies of the OECD.
[9] IPC was replaced with the Cooperative Patent Classification by USPTO and the European Patent Organization (EPO) on January 1, 2013. CPC contains new categories classified under "Y" that span different sections of the IPC in order to indicate new technological developments (Scheu *et al*., 2006; Veefkind *et al*., 2012).

A disadvantage of using USPTO data can be the relatively large changes between years in the matrix of countries versus patent classes for small economies. For example, one cannot expect a country like Slovakia to maintain a comparable patent portfolio in each year (Lengyel et al., 2015). Given this limitation in our data, we focus the discussion of empirical examples on large (e.g., the US and China) and medium-sized countries (e.g., France and Germany).

To link patents to product groups we used Eurostat ICP-NACE concordance tables (van Looy *et al*., 2015) and correspondence tables between NACE Rev.2-ISIC Rev.4; ISIC Rev.4-ISIC Rev. 3.1; ISIC Rev. 3.1-ISIC Rev. 3 (http://unstats.un.org/unsd/cr/registry/regot.asp?Lg=1), and ISIC Rev. 3-SITC Rev.3 (http://ec.europa.eu/eurostat/ramon/index.cfm?TargetUrl=DSP_PUB_WELC). However one should mention that manufacturing groups the patent-product concordance tables are available only at the 3 digit level and, furthermore, the use of more than a single table generates uncertainty since the manufacturing sectors in different classifications are not always equivalent.

MathCad is used for the mathematical derivations and SPSS (v. 23) for significance testing where appropriate.

---

[10] IPC and CPC codes are similar at the three and four-digit level.

## 4. Results

The results of the computation of ECI, PatCI, and THCI for 45 countries during the period 2000-2014 are provided in Appendices 1, 2, and 3, respectively. This information enables us to compare both the relations between and among indicators in each year, and the development of the three indicators over time.

Table 1 lists the Pearson correlations between the indicators in 2014—the last available year at the time of this research—using the full set of 45 countries. PatCI and ECI are significantly correlated ($r = .525$; $p<.01$). This correlation is not surprising given HH's argument. The correlation between THCI and ECI is stronger ($r = .774$; $p<.01$) than between THCI and PatCI ($r = .375$; $p<.01$). In sum, the pairs THCI - ECI, ECI - PatCI are correlated, but the correlation between THCI and PatCI is somewhat weaker. This may be due to large changes in patent data between years in the country-patent matrix which are enlarged by the use of Balassa's RCA: for countries with a small total number of patents which are unevenly distributed over patent classes yearly variations may lead to large changes in RCA values.

**Table 1**: Pearson correlations between the indicator values in 2014; $N$ of countries = 45.

| $N$ of countries = 45 | ECI 2014 | PatCI 2014 | THCI | ln(GDP/ population) | ln(Patents/ population) |
|---|---|---|---|---|---|
| ECI 2014 | 1 | .525[**] | .774[**] | -.094 | .077 |
| PatCI 2014 | .525[**] | 1 | .375[**] | -.241 | .000 |
| THCI | .774[**] | .375[**] | 1 | -.127 | .056 |
| ln(GDP/population) | -.094 | -.241 | -.127 | 1 | .844[**] |
| ln(Patents/population) | .077 | .000 | .056 | .844[**] | 1 |

**. Correlation is significant at the 0.01 level (2-tailed).
*. Correlation is significant at the 0.05 level (2-tailed).

We added two lines to Table 1 with the logarithms of the average income (=GDP/population) and the average number of patents per inhabitant. HH found a positive correlation between ECI and the average income, but we find a (non-significant) negative correlation. As expected, the two measures — ln(GDP/population) and ln(Patents/population) — correlate highly and significantly with each other because of the underlying correlations. (However, GDP/population can not be used on par with Patents/population, since the identical measure to Patents/population is Export value/population. Since both these measures are not correlated with the three complexity indices using our sample of the 45 relatively advanced economies, the predictive value of ECI can not be confirmed by our analysis (Hidalgo & Hausmann, 2009: Fig. 3 at p. 10573; cf. Ourens, 2013).  Our different sample choice may be one of the factors involved: Hausmann *et al.* (2011, p. 30) show that the effect of the ECI values on economic growth decreases in more developed countries. Furthermore, spurious correlations will be more prevalent in the more diversified portfolios of advanced countries. In that case, ECI would measure catching up, more than movements at the frontier of techno-economic complexity.

In order to explore this conjecture, we tested for an extended set of 119 countries at 2-digit level (SITC Rev.3) and found a remarkably higher correlation with the logarithm of GDP per capita (PPP) (Table 2). The results were also compared with ECI values obtained by HH (retrieved from http://atlas.media.mit.edu/en/rankings/country/). Both ECI values are correlated between each other, though HH's values exhibit a better correlation with logarithm GDP per capita. At the same time ECI values calculated for the set of 45 countries at 2 and 3-digit levels are not correlated with the logarithm of GDP per capita (PPP). Consequently, one can conclude that both digit levels and country choices are important parameters for the use of the ECI indicator as a predictive tool.

**Table 2**: Pearson correlations between the ECI indicator values in 2014 and GDP per capita (PPP); *N* of countries = 119, 45.

| *N* of countries = 119 | *ECI 2014 (SITC Rev.4 4-digit level)* | *ECI 2014 (SITC Rev.3 2-digit level)* | *ln(GDP/ population)* |
|---|---|---|---|
| *ECI 2014 (SITC Rev.4 4-digit level)* | 1 | .682 | .745 |
| *ECI 2014 (SITC Rev.3 2-digit level)* | .682 | 1 | .455 (n = 119) |
| *N* of countries = 45 | | | |
| *ECI 2014 (SITC Rev.3 2-digit level)* | | | .058 (n = 45) |

The dynamic evolution of the Pearson correlation between the pairs of three complexity indicators: ECI vs. PatCI, ECI vs. THCI, THCI vs. PatCI for the period 2000-2014 years is shown in Fig. 2.

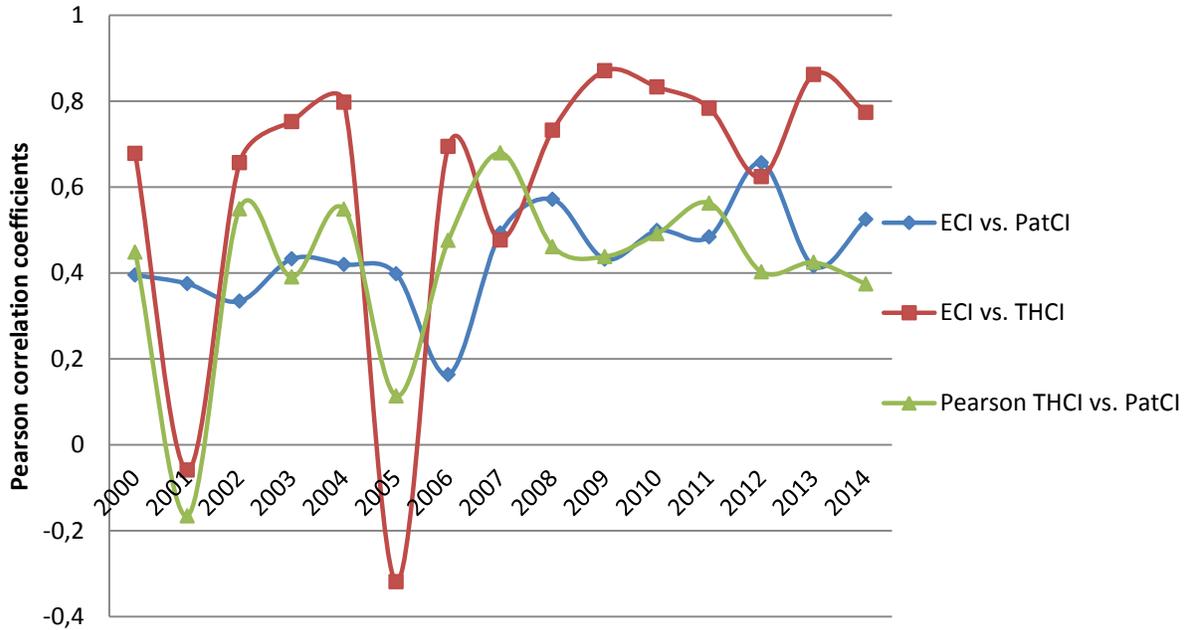

**Figure 2**: Pearson correlation between the pairs of the three complexity indicators in 2000-2014.

Two raw indicators are used to construct THCI. Consequently it is not surprising that one can expect a correlation between THCI vs. ECI and THCI vs. PatCI. Although the correlation between ECI vs. PatCI is not straightforward since formally export values and patents are not connected. The correlation reflects a link between technology and manufacturing. One can observe moderately increasing trend for the Pearson correlation between ECI and PatCI which can be explained by economic upsurge after the crisis of 1998 when the technologies were increasingly demanded by the economy. The correlations between the pairs THCI vs. ECI and THCI vs. PatCI can be considered as a coherence of dynamic interactions among the three indicators.

*The U.S.A.*

Before we turn to a comparison among countries, let us first show (Figure 3) the development of the three complexity indicators for the U.S.A. Given our use of USPTO data, one can consider PatCI to be well-defined in this case. The use of USPTO data for other countries should be considered as a proxy, and using USPTO may be an unfortunate choice for small nations (as noted above).

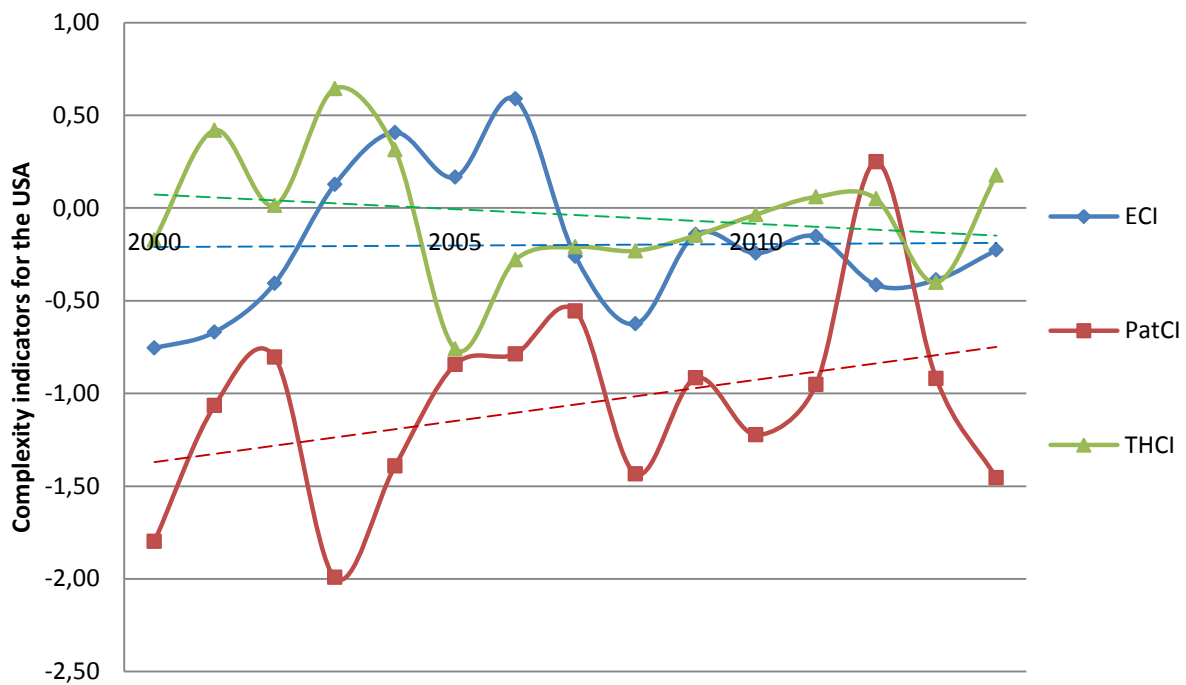

**Figure 3**: The development of the three complexity indicators in the case of the USA (2000-2014).

Figure 3 shows that PatCI values are generally below corresponding ECI and THCI values. However, the trend in PatCI is increasing. The trends for ECI and THCI are relatively stable showing oscillations. High values of THCI in comparison with ECI and PatCI indicate more interaction among the three dimensions. The figures suggest a leading role of the THCI indicator in comparison to ECI and PatCI. For example, the two peaks of THCI at 2001 and 2003 years correspond to the 2004 and 2006 peaks of ECI and the 2005 and 2007 peaks of PatCI. Furthermore, the bottom value of THCI in 2005 can be associated with ECI and PatCI bottom values in 2008.

Can THCI be perhaps considered as an early indicator? Table 3 shows positive Spearman correlations between THCI and ECI when the values of THCI are two years lagged in comparison to ECI. The value of Spearman's rho is 0.58242; $p<.05$). The association between the two variables can thus be considered statistically significant. In Figure 3, we penciled the linear regression lines in as auxiliary lines. The trends of ECI and THCI are both significant ($p < .01$), whereas the trend of PatCI is not significant.[11]

**Table 3**: Spearman correlations between complexity indicators for USA with the time shift between the THCI vs. ECI and THCI vs. PatCI indicator values of one, three, and four years.

| | ECI | | | | | PatCI | | | | |
|---|---|---|---|---|---|---|---|---|---|---|
| shift | 0 | +1 | +2 | +3 | +4 | 0 | +1 | +2 | .+3 | .+4 |
| THCI | -.011 | -.011 | .582[*] | .475 | -.027 | -.424 | .046 | -.121 | .298[*] | .114 |

    *. Correlation is significant at the 0.037 level (2-tailed).

---

[11] One can test the monotonicity of an increase or decrease for its significance using Spearman's rank-order correlation $\rho$ between the time-series data and the consecutive dates, because the latter by definition increase monotonously (Sheskin, 2011, p. 1374).

*Comparisons among countries*

Let us now compare some of the world's major economies (US, China, Japan, Germany, UK, and France). We added Russia and Canada to enrich the picture: the Canadian economy is an interesting companion to the US economy while Russia can be expected to show a different pattern.

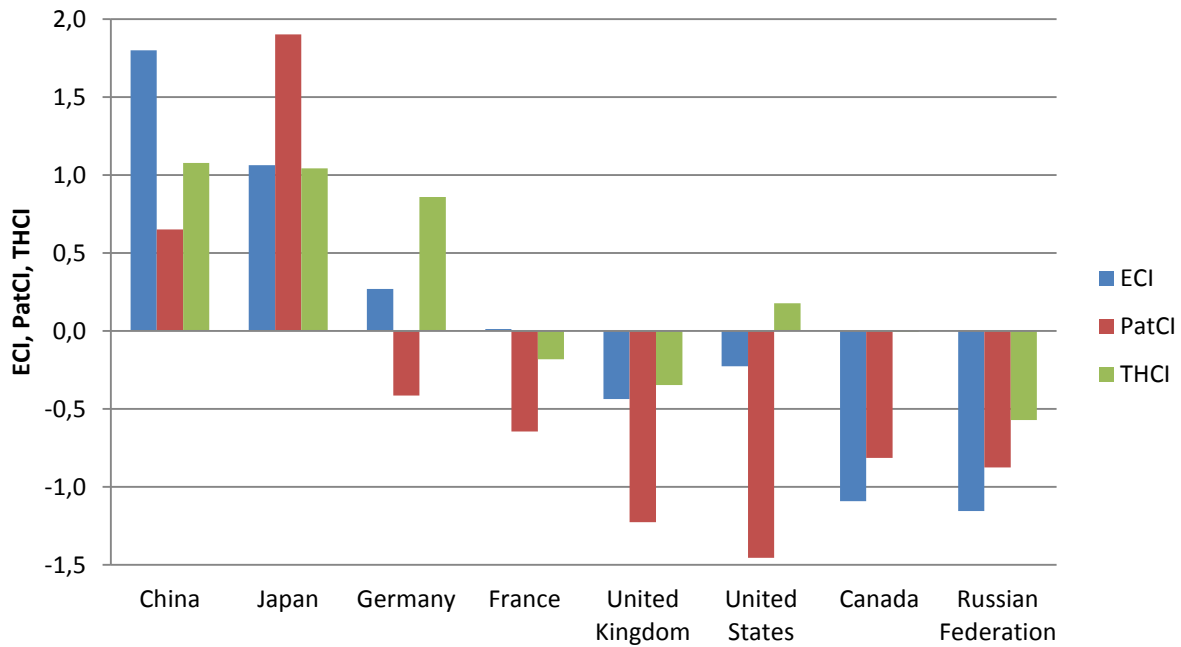

**Figure 4**: Comparison of the values of the three complexity indicators in 2014 for eight major economies (ordered by decreasing THCI).

Figure 4 first illustrates the above noted correlations between the three indicators. Values range from China and Japan with the highest positive values to the Russian Federation with the most negative values (when compared with average values). China, being the "world factory" exhibits the highest value in the group of ECI, and Japan shows the highest value of PatCI. In other words, the (non-linear) dynamics of local organization versus globalization are very different for these two countries. The US, Canada, and UK seem to share a pattern, but the economic complexity of Canada is far less than that of the UK or US. ECI is most negative for the Russian Federation, whereas France and Germany seem to form an in-between group.

The pattern of significantly lower values of ECI for Canada and the Russian Federation is confirmed by the time series of ECI for these eight countries provided in Figure 5. The most important trends to note are the statistically significant increase of China's ECI, whereas the values for both the US and UK have been significantly lower, which can be for some part ~~be~~ attributed to a decreased role of manufacturing in US and UK economies (the service sector marks up 67.8% of GDP in US[12] and 73% in UK)[13] and the effects of using Balassa's RCA. Since 2007, China scores above Japan in terms of ECI.

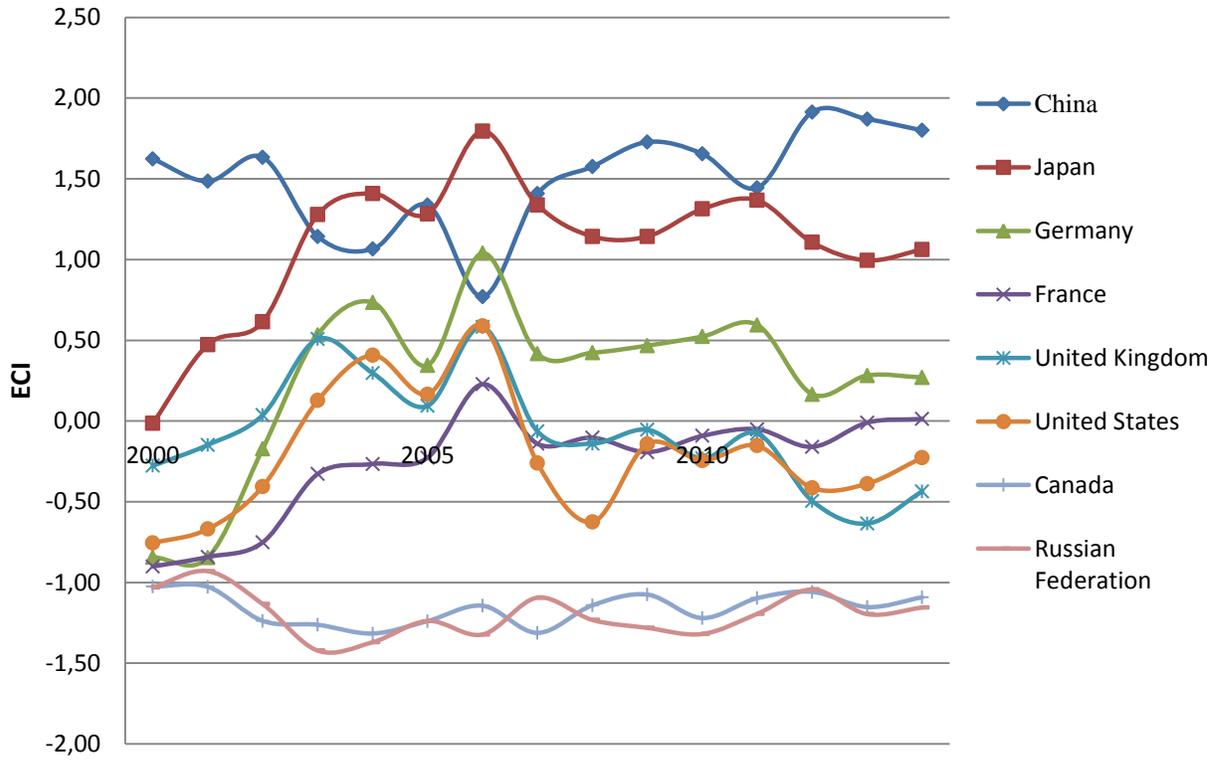

**Figure 5:** Development of ECI 2000-2014 for eight major economies.

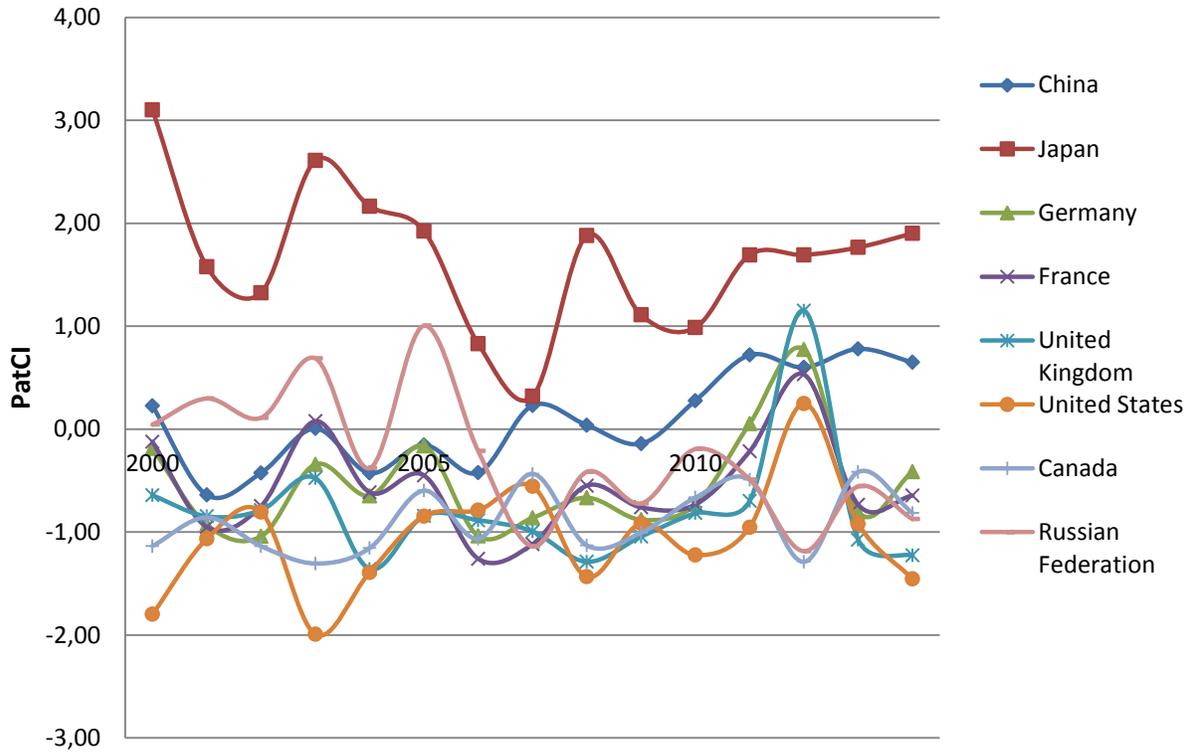

**Figure 6:** Development of PatCI 2000-2014 for eight major economies.

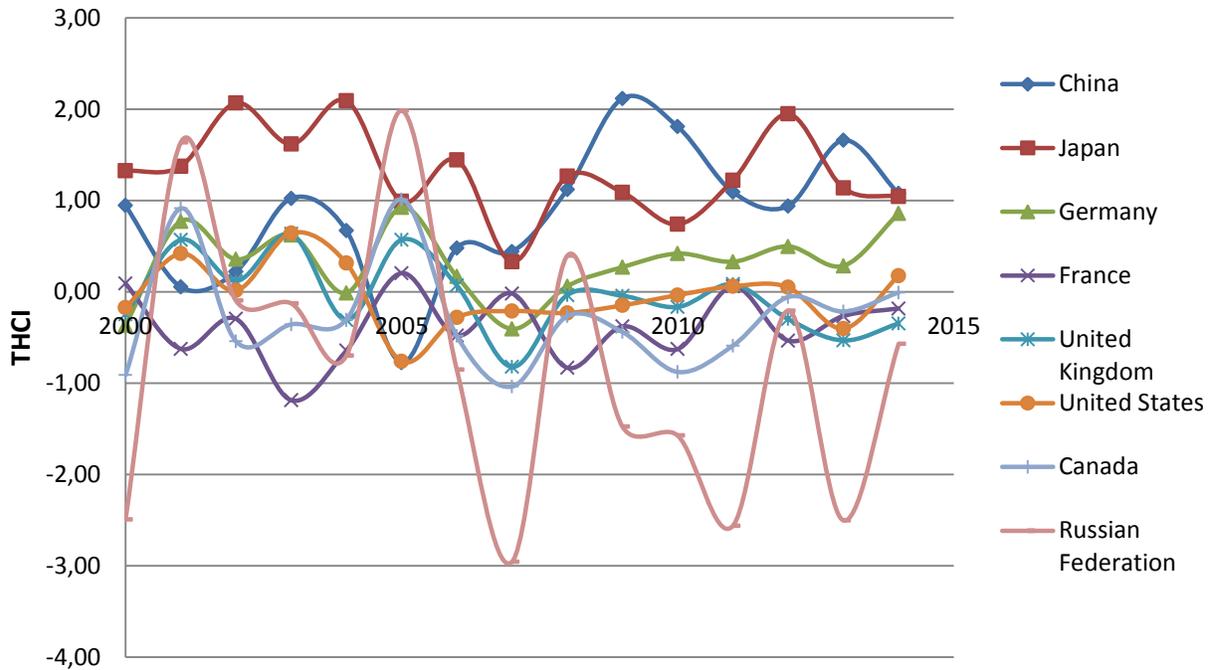

**Figure 7:** Development of THCI 2000-2014 for eight major economies.

Figure 6 shows that even for these relatively large economies, the variation in PatCI among years can be considerable. Nevertheless, Japan outperforms the other countries on this indicator in all the years under study. Figure 7 shows that Japan has a relatively stable national integration in terms of THCI, whereas for China the value of this indicator is increasing. Germany follows in the third place, whereas the Anglo-Saxon nations are relatively oriented towards the other rotation (Figures 1b and 1c); in other words, they are geared towards open markets and international patenting. The Russian Federation ranks last in the group because of a deficit in the integration at the national level (Leydesdorff, Perevodchikov, and Uvarov, 2015).

## 5. Discussion and conclusions

We argue that Hidalgo and Hausmann's Method of Reflections (MR) was only defined for economic complexity, although the authors noted that manufacturing capabilities are grounded in "capabilities" which imply knowledge-based technologies. In this study we showed how the technological dimension can be endogenized by extending MR to patent classes in analogy to product groups, and then the interaction terms can also be specified. Thus, we first defined a Patent Complexity Index (PatCI); using the country-patent matrix, ECI and PatCI were combined into the new THCI indicator. This indicator is informationally richer than either ECI or PatCI because the interaction terms can be included.

Using MR, each of the bi-lateral couplings can be considered as a helix based on the co-evolution between the respective dimensions of the corresponding matrix. The three mutually recursive formulas in the product, patent, and country dimensions can also be combined into a triangular configuration that allows for both clockwise and counter-clockwise circulation. The trade-off between the two can be captured as the Triple-Helix Complexity Index (THCI) that models the specificity of (e.g., national) systems of innovation. Systems of innovation are thus considered as the result of competition in the globalizing selection environments of markets and technologies versus (nationally organized) retention mechanisms (Freeman & Perez, 1988). In the one direction, the three helices drive one another towards systems integration; in the other direction, they lead towards differentiation and hence absorption of complexity. This model follows analytically from the MR in the complexity approach, that is, without further assumptions.

Testing the model against empirical data, we found the following results:

1. The three indicators (ECI, PatCI, THCI) are highly correlated; it is likely that all three indicators measure a latent dimension in the complexity of (national) innovation systems;

2. Our ECI values can be considered as a partial replication of HH's (2009) study albeit tested on a smaller set of 45 countries and at another (3-digit) level aggregation. Using a sample and time horizon different from HH, we were not able to reproduce the correlation between ECI and average income that has been central to the argument about the fruitfulness of the economic complexity approach. We showed that this may be caused by the fact that we only used data of a limited set of relatively developed countries and three digits instead of four digits in the product classes. A further extension of country dataset considerably improved the situation. In other words, it follows that ECI of HH is a predictor of average income at the aggregate level more than at the level of nations.

3. As to be expected, countries differ in terms of each of these indicators. Interestingly, some national systems differ across the three indicators in terms of the sign (positive or negative) despite the prevailing correlation at the level of the set. For example, France is positively assigned in terms of ECI, but negatively (with respect to the average value over the group) in terms of PatCI and THCI. A possible explanation is that French inventors patent at a below average rate in the US, whereas France's economic complexity is above average. Without repeating the analysis using data from all major patent offices, one cannot distinguish whether this is particular to France, or a limitation of the analytical procedure and the operationalization;

4. Among the major economies, Japan scores highest on all three indicators but loses ground in terms of national integration, whereas China has been increasingly successful in combining economic and technological complexity. The U.S. is assigned values below average on all three complexity indicators.

We followed HH in using available macro-level statistics. The attribution of patent classes or product groups to countries, however, requires careful elaboration (Alkemade *et al.*, 2015). When a country produces a product it may possess the corresponding technology, but the technology can also be outsourced. The manufacturing may be transferred to another country without technology transfer. The underlying dynamics among firms in terms of mergers and acquisitions are not captured when focusing on the macro-level of nations. While information about knowledge production and use should undoubtedly inform an indicator of innovation complexity, further research is required to capture the technological dimension in a meaningful way. For example, one would have to define a measure which plays the role of GDP/population in the technological domain.

6. **Normative implications**

The search for indicators of an economy's knowledge base that are, on the one hand, not external to the economy (as are most science & technology indicators) yet on the other hand, endogenize technological developments as capacities has been a major driver of the research program of the OECD (David & Foray, 1995; Foray & Lundvall, 1996). In an evaluation of these efforts, Godin (2006) concluded that they mainly had led to a re-organization and relabeling of existing indicators.

Within the context of Triple-Helix (TH) research, mutual information among three (or more) dimensions has been developed as an indicator of synergy and integration in the economy's knowledge base (e.g., Leydesdorff & Fritsch, 2006). However, this synergy indicator has several problems, such as: (1) it is measured in bits of information, which is an abstract entropy measure and cannot be directly related to basic economic measures (such as turnover, income, etc.); and (2) as a systems measure, synergy cannot be decomposed proportionally among the three sub-dynamics—technological trajectories, market selections, and control mechanisms—governing the TH's evolution. In this study, we propose a solution of this problem by elaborating the complexity approach in two and three dimensions.

We showed that ECI can analytically be elaborated into a Triple Helix Complexity Indicator of innovation systems. Thus, it is possible to generalize the Method of Reflections to include the interactions among the three relevant dimensions: countries, products, and technologies. Endogenization of the technological dimension could mean that this methodology would have a better potential to accurately predict a country's economic growth. In this sense, the current study can be considered as a contribution to the longer-term effort to study complexity as a measure of innovation systems and the knowledge base of an economy (Frenken *et al.*, 2007; Ivanova & Leydesdorff, 2014).

Given the state of the art, it is perhaps too early for the formulation of normative advice. One could explore first whether the indicators can be improved and made more meaningful. For example, the choice of RCA (Balassa, 1965) as a binary threshold implies throwing away a lot of information in the valued (!) data. Petersen *et al.* (2016) found that their information-theoretical indicator of synergy in innovation systems did no longer work after a reduction to a binary reflection. Valued variants of ECI, PatCI, and THCI may provide us with stronger predictors.

## Acknowledgements

We thank Koen Frenken and Frank Neffke for valuable comments on an earlier version of this paper. Inga Ivanova acknowledges financial support from the Basic Research Program implemented at the National Research University Higher School of Economics and the Russian Academic Excellence Project '5-100'. We are grateful to Imogen Wade for editing and proofreading.

Appendix 1
Economic Complexity Index (ECI) for 45 countries for the period 2000-2014

| | 2000 | 2001 | 2002 | 2003 | 2004 | 2005 | 2006 | 2007 | 2008 | 2009 | 2010 | 2011 | 2012 | 2013 | 2014 |
|---|---|---|---|---|---|---|---|---|---|---|---|---|---|---|---|
| Argentina | -0.71 | -0.63 | -1.19 | -1.49 | -1.54 | -1.33 | -1.44 | -1.44 | -1.44 | -1.34 | -1.15 | -1.29 | -1.19 | -1.14 | -1.37 |
| Austria | -0.60 | -0.62 | 0.06 | 0.59 | 0.59 | 0.33 | 0.81 | 0.56 | 0.73 | 0.61 | 0.64 | 0.77 | 0.55 | 0.66 | 0.52 |
| Australia | -1.08 | -1.05 | -1.82 | -1.95 | -2.10 | -2.09 | -2.27 | -2.09 | -2.34 | -2.33 | -2.17 | -2.40 | -2.32 | -2.38 | -2.32 |
| Belgium | -0.53 | -0.45 | -0.47 | -0.37 | -0.43 | -0.44 | -0.26 | -0.54 | -0.60 | -0.35 | -0.52 | -0.42 | -0.58 | -0.51 | -0.41 |
| Brazil | -0.46 | -0.37 | -0.88 | -0.99 | -1.11 | -0.83 | -0.89 | -0.82 | -1.06 | -0.87 | -0.87 | -1.05 | -0.93 | -1.02 | -0.84 |
| Canada | -1.03 | -1.03 | -1.24 | -1.26 | -1.32 | -1.24 | -1.14 | -1.14 | -1.14 | -1.08 | -1.22 | -1.10 | -1.06 | -1.15 | -1.09 |
| Switzerland | -0.48 | -0.56 | 0.12 | 0.85 | 0.97 | 0.71 | 1.23 | 0.64 | 0.20 | 0.30 | 0.21 | 0.12 | -0.18 | 0.04 | 0.06 |
| Chile | -1.05 | -0.96 | -1.67 | -2.29 | -2.35 | -2.24 | -2.32 | -2.21 | -2.02 | -2.08 | -2.21 | -2.15 | -1.84 | -1.84 | -1.87 |
| China | 1.62 | 1.49 | 1.63 | 1.14 | 1.07 | 1.34 | 0.77 | 1.41 | 1.58 | 1.73 | 1.66 | 1.44 | 1.91 | 1.87 | 1.80 |
| Czech Republic | -0.09 | -0.15 | 0.44 | 0.70 | 0.66 | 0.72 | 0.86 | 0.85 | 1.00 | 1.15 | 1.11 | 1.20 | 0.77 | 0.89 | 0.84 |
| Germany | -0.84 | -0.85 | -0.17 | 0.53 | 0.74 | 0.35 | 1.04 | 0.42 | 0.42 | 0.47 | 0.52 | 0.60 | 0.17 | 0.28 | 0.27 |
| Denmark | -0.43 | -0.66 | -0.33 | -0.03 | -0.06 | -0.30 | -0.04 | -0.18 | 0.00 | -0.05 | 0.06 | 0.09 | 0.02 | 0.10 | -0.03 |
| Estonia | 0.17 | -0.02 | 0.04 | -0.20 | -0.03 | 0.02 | -0.31 | -0.16 | 0.13 | 0.22 | 0.33 | 0.37 | 0.35 | 0.36 | 0.23 |
| Spain | -0.09 | -0.17 | 0.06 | 0.02 | -0.16 | -0.33 | -0.32 | -0.18 | 0.21 | -0.19 | -0.23 | -0.07 | -0.07 | 0.17 | 0.02 |
| Finland | -0.90 | -0.92 | -0.83 | -0.56 | -0.47 | -0.27 | 0.05 | -0.22 | -0.02 | -0.09 | 0.02 | 0.02 | -0.22 | -0.20 | -0.16 |
| France | -0.90 | -0.84 | -0.75 | -0.33 | -0.27 | -0.23 | 0.23 | -0.14 | -0.10 | -0.19 | -0.09 | -0.05 | -0.16 | -0.01 | 0.01 |
| United Kingdom | -0.28 | -0.15 | 0.04 | 0.51 | 0.30 | 0.09 | 0.58 | -0.06 | -0.14 | -0.05 | -0.23 | -0.08 | -0.50 | -0.64 | -0.44 |
| Greece | 0.27 | 0.24 | 0.26 | 0.10 | -0.05 | -0.08 | -0.39 | 0.03 | 0.18 | -0.10 | -0.08 | -0.31 | -0.29 | -0.39 | -0.42 |
| Hong Kong | 2.75 | 2.67 | 2.70 | 2.37 | 2.25 | 2.46 | 1.58 | 2.33 | 1.73 | 2.44 | 2.19 | 1.84 | 2.56 | 2.40 | 2.48 |
| Hungary | 0.69 | 0.72 | 0.74 | 0.59 | 0.43 | 0.33 | 0.78 | 0.56 | 0.62 | 0.52 | 0.67 | 0.69 | 0.53 | 0.59 | 0.68 |
| Indonesia | 1.49 | 1.23 | 0.82 | 0.07 | 0.11 | 0.38 | -0.46 | 0.136 | 0.18 | -0.07 | 0.06 | -0.12 | 0.23 | 0.03 | 0.18 |
| Ireland | -0.43 | -0.55 | -0.65 | -0.19 | -0.18 | -0.19 | -0.09 | -0.49 | -1.21 | -0.79 | -0.66 | -1.01 | -1.12 | -0.98 | -1.08 |
| Israel | 0.28 | 0.62 | 0.49 | 0.48 | 0.89 | 0.33 | 0.12 | -0.35 | -0.42 | -0.14 | -0.27 | -0.17 | -0.32 | -0.29 | -0.34 |
| India | 0.85 | 0.97 | 0.59 | 0.28 | 0.10 | 0.54 | -0.26 | 0.06 | 0.23 | 0.07 | 0.11 | 0.17 | 0.44 | 0.47 | 0.55 |
| Iceland | -0.33 | -1.00 | -1.20 | -1.68 | -1.12 | -1.65 | -1.87 | -1.74 | -1.98 | -2.06 | -2.30 | -2.28 | -1.80 | -2.08 | -2.18 |
| Italy | -0.01 | -0.01 | 0.45 | 0.78 | 0.75 | 0.74 | 0.83 | 0.89 | 0.99 | 0.88 | 0.91 | 0.96 | 1.01 | 1.02 | 0.96 |
| Japan | -0.01 | 0.47 | 0.62 | 1.28 | 1.41 | 1.28 | 1.80 | 1.34 | 1.14 | 1.14 | 1.31 | 1.37 | 1.11 | 1.00 | 1.06 |
| Korea, Republic of | 1.11 | 1.39 | 1.27 | 1.30 | 1.10 | 1.61 | 1.42 | 1.52 | 1.29 | 1.37 | 1.44 | 1.45 | 1.55 | 1.33 | 1.27 |
| Luxembourg | -0.63 | -0.52 | -0.32 | 0.01 | -0.04 | 0.03 | 0.24 | 0.30 | 0.65 | 0.40 | 0.37 | 0.25 | 0.43 | 0.40 | 0.54 |
| Mexico | 0.87 | 0.96 | 1.28 | 1.15 | 0.99 | 0.93 | 0.68 | 0.66 | 0.55 | 0.67 | 0.59 | 0.62 | 0.78 | 0.83 | 0.95 |
| Malaysia | 2.85 | 2.66 | 1.89 | 1.03 | 1.05 | 1.42 | 0.68 | 1.28 | 0.79 | 0.97 | 1.04 | 0.78 | 0.99 | 0.48 | 0.71 |
| Netherlands | -0.36 | -0.26 | -0.70 | -0.69 | -0.59 | -0.49 | -0.46 | -0.55 | -0.63 | -0.55 | -0.52 | -0.64 | -0.60 | -0.63 | -0.60 |
| Norway | -1.01 | -1.16 | -1.09 | -0.96 | -0.91 | -0.94 | -0.54 | -0.75 | -0.54 | -0.86 | -0.95 | -0.92 | -0.91 | -0.89 | -0.92 |
| New Zealand | -1.09 | -1.05 | -1.61 | -1.69 | -1.78 | -1.73 | -1.70 | -1.51 | -1.54 | -1.47 | -1.43 | -1.39 | -1.46 | -1.27 | -1.40 |
| Poland | -0.29 | -0.37 | -0.06 | -0.10 | -0.32 | -0.29 | -0.19 | -0.12 | 0.14 | 0.06 | 0.06 | 0.19 | 0.06 | 0.20 | 0.05 |
| Portugal | 0.68 | 0.56 | 0.62 | 0.41 | 0.31 | 0.31 | -0.03 | 0.34 | 0.54 | 0.44 | 0.37 | 0.51 | 0.69 | 0.75 | 0.66 |
| Romania | 0.45 | 0.53 | 0.73 | 0.39 | 0.50 | 0.60 | 0.40 | 1.09 | 1.33 | 1.15 | 0.90 | 1.03 | 0.98 | 1.10 | 1.10 |
| Russian Federation | -1.03 | -0.93 | -1.13 | -1.42 | -1.37 | -1.24 | -1.32 | -1.09 | -1.23 | -1.28 | -1.32 | -1.19 | -1.04 | -1.20 | -1.16 |
| Sweden | -0.82 | -0.88 | -0.48 | 0.05 | 0.01 | -0.22 | 0.21 | 0.09 | 0.12 | -0.07 | 0.02 | 0.36 | 0.12 | 0.11 | 0.01 |
| Singapore | 2.59 | 2.59 | 1.82 | 1.42 | 1.65 | 1.47 | 1.54 | 1.43 | 0.84 | 0.91 | 0.68 | 0.36 | 0.41 | 0.35 | 0.33 |
| Slovenia | -0.17 | -0.23 | 0.28 | 0.72 | 0.69 | 0.43 | 0.66 | 0.42 | 0.81 | 0.38 | 0.51 | 0.73 | 0.57 | 0.75 | 0.64 |
| Slovakia | -0.13 | -0.24 | 0.20 | 0.24 | 0.21 | 0.38 | 0.59 | 0.74 | 0.94 | 0.58 | 0.92 | 0.99 | 0.71 | 0.85 | 0.87 |
| Turkey | 0.55 | 0.81 | 0.87 | 0.56 | 0.48 | 0.54 | 0.13 | 0.58 | 0.98 | 0.82 | 0.86 | 0.91 | 1.10 | 1.06 | 0.95 |
| United States | -0.76 | -0.67 | -0.41 | 0.13 | 0.41 | 0.17 | 0.59 | -0.26 | -0.63 | -0.14 | -0.24 | -0.15 | -0.42 | -0.39 | -0.23 |
| South Africa | -0.70 | -0.63 | -1.06 | -1.48 | -1.50 | -1.37 | -1.50 | -1.46 | -1.30 | -1.13 | -1.08 | -1.04 | -1.03 | -1.08 | -0.87 |

Appendix 2
Patent Complexity Index (PatCI) for 45 countries for the period 2000-2014

| | 2000 | 2001 | 2002 | 2003 | 2004 | 2005 | 2006 | 2007 | 2008 | 2009 | 2010 | 2011 | 2012 | 2013 | 2014 |
|---|---|---|---|---|---|---|---|---|---|---|---|---|---|---|---|
| Argentina | -0.48 | -0.02 | -0.51 | -0.56 | 0.21 | -0.71 | -0.10 | -0.46 | -0.41 | 0.25 | 0.17 | -0.68 | -0.61 | -0.57 | -0.67 |
| Austria | -0.64 | -1.04 | -0.86 | -0.70 | -0.61 | -0.21 | -0.76 | -0.51 | -0.44 | -0.57 | -0.76 | -0.21 | 0.30 | -0.34 | -0.29 |
| Australia | -1.20 | -1.23 | -0.99 | -1.70 | -1.09 | -0.59 | -0.55 | -1.15 | -1.40 | -1.24 | -1.60 | -0.51 | -2.54 | -1.35 | -1.24 |
| Belgium | 0.39 | -0.91 | -0.41 | 0.33 | -0.46 | -0.24 | -0.80 | -1.12 | -0.50 | -1.00 | -0.64 | -0.19 | -0.40 | -0.72 | -0.53 |
| Brazil | -0.42 | -0.50 | -0.18 | -0.26 | -0.28 | -0.56 | -0.98 | -0.61 | -0.65 | -0.45 | -0.61 | -0.63 | -1.43 | -0.84 | -0.86 |
| Canada | -1.14 | -0.86 | -1.14 | -1.31 | -1.16 | -0.60 | -1.06 | -0.43 | -1.13 | -1.01 | -0.67 | -0.49 | -1.29 | -0.42 | -0.82 |
| Switzerland | -0.44 | -0.72 | -0.77 | -0.36 | -0.35 | -0.20 | -0.88 | -0.87 | -0.45 | -0.96 | -1.00 | -0.39 | 1.79 | -0.94 | -0.74 |
| Chile | -0.23 | 0.21 | -0.41 | -0.56 | -1.42 | -0.43 | -0.60 | -0.86 | -0.35 | -0.17 | -1.36 | -0.88 | -1.53 | -0.66 | -0.98 |
| China | 0.23 | -0.64 | -0.43 | 0.01 | -0.43 | -0.16 | -0.42 | 0.23 | 0.04 | -0.14 | 0.28 | 0.72 | 0.60 | 0.78 | 0.65 |
| Czech Republic | 0.47 | 0.70 | 0.61 | 1.52 | 1.14 | 0.94 | 0.09 | 0.50 | 0.77 | 0.24 | 0.87 | 0.22 | 0.16 | 0.03 | 0.50 |
| Germany | -0.19 | -0.93 | -1.04 | -0.34 | -0.65 | -0.16 | -1.04 | -0.86 | -0.67 | -0.88 | -0.73 | 0.05 | 0.77 | -0.81 | -0.41 |
| Denmark | -0.54 | -1.12 | -1.37 | -1.55 | -0.64 | -0.54 | -0.98 | -0.54 | -0.87 | -0.41 | -0.77 | -0.28 | -0.12 | -1.01 | -0.76 |
| Estonia | 2.05 | 1.58 | 3.12 | 0.33 | -0.49 | -0.52 | 1.43 | 0.77 | 0.80 | 0.35 | 0.99 | -0.37 | -0.66 | 0.47 | 1.15 |
| Spain | -0.87 | -1.22 | -0.82 | -1.22 | -0.90 | -0.66 | -0.95 | -0.52 | -0.73 | -0.62 | -0.99 | -0.52 | -0.08 | -0.64 | -0.86 |
| Finland | -0.55 | -1.08 | -0.91 | -0.68 | -0.99 | -0.39 | -1.33 | -0.44 | -0.93 | -1.38 | -1.09 | 0.14 | -0.60 | 0.33 | 0.63 |
| France | -0.12 | -0.95 | -0.75 | 0.08 | -0.61 | -0.45 | -1.26 | -1.12 | -0.55 | -0.76 | -0.73 | -0.22 | 0.53 | -0.74 | -0.65 |
| United Kingdom | -0.64 | -0.85 | -0.79 | -0.48 | -1.36 | -0.84 | -0.89 | -0.99 | -1.29 | -1.04 | -0.82 | -0.70 | 1.15 | -1.08 | -1.23 |
| Greece | -1.01 | -0.01 | 0.70 | 0.08 | 0.01 | -0.85 | 1.35 | 0.33 | 0.85 | 0.51 | 0.24 | -0.16 | -0.80 | 0.25 | -0.05 |
| Hong Kong | -0.28 | -0.44 | -0.10 | -0.24 | 0.22 | -0.01 | 0.33 | 0.76 | -0.11 | 0.94 | -0.26 | 0.54 | 0.75 | -0.33 | 0.07 |
| Hungary | 0.32 | 0.49 | 0.93 | 1.29 | 1.21 | -0.50 | -0.22 | 1.27 | 0.79 | 0.42 | 0.56 | -0.14 | 0.27 | -0.05 | 0.17 |
| Indonesia | -0.09 | 0.70 | -0.58 | -0.04 | 2.06 | -0.54 | 1.35 | -0.32 | 0.23 | 2.16 | 1.40 | 1.29 | -1.13 | -0.02 | 0.33 |
| Ireland | -0.40 | 0.70 | 0.74 | 0.15 | -0.08 | 0.06 | 1.27 | 0.48 | -0.42 | -0.21 | 0.48 | -0.26 | 0.88 | 0.33 | 0.14 |
| Israel | -0.82 | 0.40 | 0.30 | 0.13 | -0.83 | -0.44 | 0.05 | -0.24 | -0.32 | -0.26 | 0.41 | -0.03 | 1.06 | 1.46 | -0.31 |
| India | -0.10 | 0.36 | -0.22 | 0.20 | -0.30 | -0.82 | -0.25 | -0.56 | -0.09 | -0.41 | 0.66 | 0.48 | 0.53 | 1.73 | 1.21 |
| Iceland | -0.79 | 2.51 | 1.66 | 0.70 | 1.12 | -1.07 | 1.19 | 0.50 | -0.33 | 0.35 | 0.16 | -1.06 | -2.01 | 0.23 | 0.57 |
| Italy | -0.32 | -0.80 | -0.91 | -0.78 | -0.57 | -0.42 | -1.01 | -0.54 | -0.50 | -0.81 | -0.78 | -0.49 | 0.76 | -0.63 | -0.53 |
| Japan | 3.10 | 1.58 | 1.33 | 2.61 | 2.17 | 1.92 | 0.83 | 0.32 | 1.88 | 1.11 | 0.99 | 1.69 | 1.69 | 1.77 | 1.90 |
| Korea, Republic of | 2.15 | 1.52 | 1.11 | 2.08 | 0.83 | 1.79 | 1.39 | 1.62 | 2.50 | 3.67 | 2.63 | 4.75 | 1.74 | 3.16 | 3.72 |
| Luxembourg | -0.45 | -0.26 | 0.02 | -0.29 | 0.17 | 0.24 | -0.34 | -1.00 | 0.47 | -0.58 | 0.54 | -0.77 | 0.46 | -0.95 | 0.19 |
| Mexico | -0.62 | -0.51 | -1.01 | -0.44 | -0.09 | 0.43 | -0.71 | -0.10 | -0.12 | -0.33 | -0.34 | -0.45 | 0.69 | -0.33 | -0.07 |
| Malaysia | 1.46 | 0.97 | 1.24 | 1.30 | 0.58 | 1.71 | 1.15 | 2.02 | 1.84 | 1.18 | 0.80 | 1.01 | -0.15 | 0.71 | 1.39 |
| Netherlands | 0.20 | -0.65 | -0.42 | 0.54 | -0.26 | -0.04 | -0.13 | -0.50 | 0.21 | -0.13 | 0.17 | 1.10 | 0.19 | 0.14 | -0.53 |
| Norway | -0.65 | -1.05 | -1.12 | -1.49 | -0.78 | -0.19 | -0.66 | -0.61 | -0.99 | -0.80 | -1.32 | -0.75 | -1.31 | -0.21 | -0.92 |
| New Zealand | -0.29 | -0.81 | -0.64 | -0.73 | -0.60 | -0.41 | -0.53 | -0.22 | -0.60 | 0.10 | -0.67 | -0.75 | -0.97 | -0.81 | -0.66 |
| Poland | 0.91 | 0.67 | 1.03 | 0.21 | 0.62 | 0.49 | 0.13 | 0.34 | -0.49 | -0.33 | 0.95 | 0.11 | 0.06 | -0.02 | 0.14 |
| Portugal | 0.43 | 1.63 | 0.28 | 1.32 | 1.89 | -1.06 | 1.01 | 1.64 | 1.90 | 0.95 | 0.12 | -0.56 | 0.04 | 0.07 | 1.78 |
| Romania | 1.78 | 1.77 | 1.76 | 0.73 | 2.20 | 0.70 | 3.24 | 3.37 | 2.25 | 1.78 | 3.10 | 1.38 | -0.01 | 2.97 | 0.69 |
| Russian Federation | 0.04 | 0.30 | 0.11 | 0.69 | -0.38 | 1.00 | -0.21 | -1.14 | -0.42 | -0.72 | -0.19 | -0.49 | -1.19 | -0.56 | -0.88 |
| Sweden | -0.70 | -0.81 | -0.80 | -0.91 | -1.10 | -0.17 | -0.65 | -0.84 | -0.76 | -0.47 | -0.65 | -0.45 | -1.01 | -0.68 | -0.66 |
| Singapore | 0.59 | 0.53 | 1.44 | 1.82 | 1.53 | 1.96 | 1.07 | 1.03 | 0.53 | 0.62 | 0.88 | 1.61 | 1.38 | 1.68 | 1.41 |
| Slovenia | -0.64 | 0.35 | 1.53 | 0.80 | 1.34 | -0.52 | 1.05 | 2.34 | 1.03 | 1.73 | 0.66 | 0.33 | 0.35 | -0.56 | 0.57 |
| Slovakia | 0.54 | 1.22 | 1.02 | 0.51 | 0.59 | 4.32 | 1.53 | -0.18 | -0.52 | 0.61 | 0.36 | -0.54 | 0.11 | 0.54 | 0.11 |
| Turkey | 2.30 | 1.43 | -0.29 | 0.17 | 0.88 | -0.18 | 0.29 | 0.20 | 2.00 | 0.54 | 1.05 | -0.90 | 0.59 | 0.07 | -0.11 |
| United States | -1.80 | -1.07 | -0.81 | -1.99 | -1.39 | -0.85 | -0.79 | -0.56 | -1.43 | -0.92 | -1.22 | -0.95 | 0.25 | -0.92 | -1.46 |
| South Africa | -0.52 | -1.15 | -0.68 | -0.99 | -0.94 | -0.26 | -0.67 | -0.41 | -0.65 | -0.90 | -1.25 | -0.61 | -0.79 | -0.55 | -1.16 |

Appendix 3
Triple Helix Complexity Index (THCI) for 45 countries for the period 2000-2014.

| | 2000 | 2001 | 2002 | 2003 | 2004 | 2005 | 2006 | 2007 | 2008 | 2009 | 2010 | 2011 | 2012 | 2013 | 2014 |
|---|---|---|---|---|---|---|---|---|---|---|---|---|---|---|---|
| Argentina | -0.21 | -0.83 | -1.52 | -1.81 | -1.93 | -1.45 | -1.91 | -0.34 | -2.02 | -1.47 | -0.92 | -1.60 | -1.47 | -0.73 | -1.99 |
| Austria | 0.03 | 0.52 | 0.25 | 0.49 | 0.30 | -0.47 | 0.65 | -0.25 | 0.89 | 0.44 | 0.31 | 0.64 | 1.10 | 0.90 | 0.82 |
| Australia | -1.84 | 0.14 | -1.94 | -0.88 | -1.92 | 1.38 | -1.54 | -1.54 | -0.05 | -2.16 | -0.81 | -1.50 | -0.26 | -2.18 | -0.99 |
| Belgium | -0.99 | -0.40 | -1.04 | -1.34 | -1.46 | -0.31 | -1.32 | -0.15 | -1.09 | -1.15 | -1.07 | -1.10 | -1.17 | -1.58 | -0.81 |
| Brazil | -1.15 | 0.21 | -0.48 | -0.58 | -1.07 | 0.76 | -0.64 | -1.00 | -0.66 | -0.93 | -0.69 | -0.78 | -0.08 | -0.87 | -0.53 |
| Canada | -0.91 | 0.91 | -0.54 | -0.36 | -0.30 | 1.01 | -0.48 | -1.04 | -0.27 | -0.44 | -0.88 | -0.59 | -0.06 | -0.22 | -0.01 |
| Switzerland | 0.00 | 0.27 | 0.10 | 0.46 | 0.58 | -0.47 | -0.04 | -0.41 | 0.24 | 0.22 | 0.43 | 0.32 | 0.21 | 0.15 | 0.22 |
| Chile | -1.08 | -1.15 | -1.42 | -2.44 | -1.69 | 0.55 | -2.35 | -1.06 | -1.79 | -1.03 | -2.41 | -1.94 | -1.16 | -1.60 | -1.51 |
| China | 0.94 | 0.06 | 0.22 | 1.02 | 0.67 | -0.78 | 0.48 | 0.44 | 1.12 | 2.12 | 1.81 | 1.09 | 0.94 | 1.66 | 1.08 |
| Czech Republic | 0.53 | 0.48 | 0.75 | 1.48 | 0.66 | -0.48 | 0.98 | 0.81 | 0.51 | 0.97 | 0.84 | 0.32 | 1.52 | 0.87 | 1.05 |
| Germany | -0.37 | 0.77 | 0.36 | 0.62 | -0.01 | 0.93 | 0.17 | -0.41 | 0.06 | 0.27 | 0.42 | 0.33 | 0.50 | 0.29 | 0.86 |
| Denmark | -0.23 | -0.90 | -1.01 | -0.55 | 0.09 | -1.08 | -0.39 | 0.89 | -0.81 | 0.11 | -0.23 | 0.03 | -0.65 | -0.07 | 0.12 |
| Estonia | 0.98 | 0.41 | 2.15 | -0.35 | 0.75 | -0.75 | -1.19 | 2.01 | 0.19 | -1.08 | -0.77 | 0.24 | 0.04 | 1.69 | 0.28 |
| Spain | 0.35 | -0.66 | -0.61 | -0.56 | -0.87 | -0.99 | -0.61 | 0.25 | -0.75 | -0.58 | -0.37 | -0.59 | -0.42 | 0.19 | -0.36 |
| Finland | -1.01 | 1.41 | 0.40 | 0.60 | 0.41 | 1.84 | 0.23 | -0.62 | 0.97 | 0.52 | 0.60 | 1.41 | 1.18 | -0.16 | 0.88 |
| France | 0.09 | -0.62 | -0.30 | -1.19 | -0.64 | 0.21 | -0.48 | -0.02 | -0.83 | -0.38 | -0.63 | 0.06 | -0.53 | -0.27 | -0.18 |
| United Kingdom | -0.25 | 0.57 | 0.12 | 0.63 | -0.31 | 0.57 | 0.07 | -0.82 | -0.04 | -0.04 | -0.17 | 0.08 | -0.30 | -0.53 | -0.35 |
| Greece | 0.29 | -1.11 | -1.05 | -0.76 | -1.04 | 0.99 | 0.41 | 0.79 | 0.00 | 0.04 | -0.14 | 0.15 | -1.45 | -0.68 | -1.59 |
| Hong Kong | 1.78 | 0.18 | 1.10 | 2.09 | 1.20 | -2.10 | 1.22 | 1.12 | 1.75 | 2.43 | 2.74 | 2.42 | 1.03 | 2.59 | 1.77 |
| Hungary | 0.09 | 0.00 | 0.02 | 0.48 | 1.03 | 0.21 | 0.51 | 0.84 | -0.33 | 0.53 | -0.27 | -0.47 | -0.47 | 0.52 | -0.68 |
| Indonesia | 0.05 | -1.58 | -0.39 | 0.32 | 1.27 | -1.11 | 2.17 | -0.65 | -0.75 | -0.02 | 0.79 | 0.39 | -0.19 | -0.46 | -0.15 |
| Ireland | 0.58 | -0.96 | -0.68 | -0.18 | -0.92 | 0.10 | -0.99 | 1.70 | -1.00 | -1.20 | 0.18 | -1.41 | -1.30 | -0.88 | -1.07 |
| Israel | 0.83 | 0.25 | 1.04 | 1.63 | 1.36 | -0.38 | -0.01 | 0.29 | 0.17 | 0.57 | 0.47 | 0.74 | 0.16 | 0.24 | -0.38 |
| India | -0.69 | -1.65 | -0.68 | -0.74 | -1.14 | -0.84 | -0.91 | -0.30 | -0.98 | -0.81 | 0.04 | 0.04 | -0.98 | 0.13 | -0.72 |
| Iceland | 1.73 | -1.49 | -0.78 | -0.07 | -0.41 | -0.24 | 0.01 | 0.41 | -2.04 | -1.41 | -1.47 | -1.19 | -2.10 | -1.24 | -1.25 |
| Italy | 0.96 | -0.44 | 0.08 | 0.19 | -0.28 | -1.02 | 0.43 | 0.21 | 0.07 | 0.65 | 0.60 | 0.19 | 0.41 | 0.77 | 0.75 |
| Japan | 1.32 | 1.38 | 2.07 | 1.62 | 2.09 | 0.99 | 1.45 | 0.33 | 1.27 | 1.09 | 0.74 | 1.22 | 1.95 | 1.14 | 1.04 |
| Korea, Republic of | 1.31 | 0.78 | 1.83 | 1.77 | 1.23 | 0.65 | 1.41 | 1.01 | 1.58 | 1.64 | 1.74 | 1.95 | 1.39 | 0.92 | 1.06 |
| Luxembourg | -0.58 | 0.79 | 0.13 | -0.09 | 0.07 | 0.60 | 0.53 | -0.27 | 0.40 | 0.15 | -0.15 | -0.28 | 0.48 | -0.01 | 1.11 |
| Mexico | -0.19 | -0.21 | 0.40 | 0.21 | 0.52 | -0.70 | 0.53 | 0.23 | 0.26 | -0.03 | -0.01 | -0.10 | 0.63 | 0.26 | 0.70 |
| Malaysia | 2.04 | 0.12 | 1.15 | 0.34 | 1.09 | -0.19 | 0.53 | 0.77 | 1.09 | 0.21 | 0.47 | -0.67 | 0.55 | 0.39 | 0.37 |
| Netherlands | -0.34 | -0.76 | -0.93 | -1.47 | -0.86 | -0.07 | -1.05 | -0.37 | -0.92 | -0.80 | -0.58 | -0.33 | -1.28 | -0.74 | -1.46 |
| Norway | -2.09 | 1.00 | -0.38 | -0.25 | -1.17 | 1.22 | -0.25 | -1.36 | 0.22 | -0.48 | -0.53 | -0.38 | 0.51 | -0.74 | -0.23 |
| New Zealand | -0.18 | -0.44 | -1.89 | -1.95 | -1.33 | -0.72 | -1.62 | 0.39 | -1.76 | -0.75 | -0.57 | -0.30 | -1.01 | 0.01 | -0.50 |
| Poland | -0.26 | -0.12 | -0.40 | -0.81 | 0.71 | -0.89 | -0.25 | 0.87 | -0.36 | 0.01 | -0.68 | 0.03 | 0.05 | 0.19 | -0.46 |
| Portugal | 1.04 | -0.47 | 0.59 | 0.35 | 1.09 | 0.00 | 1.36 | 1.82 | -0.69 | -0.43 | -0.15 | 0.26 | -1.25 | 0.61 | 0.05 |
| Romania | 0.16 | 1.20 | 2.51 | 1.09 | 0.22 | 0.66 | 2.08 | 1.73 | 2.18 | 2.00 | 1.95 | 1.45 | 1.04 | 1.83 | 1.38 |
| Russian Federation | -2.49 | 1.63 | -0.10 | -0.13 | -0.70 | 1.98 | -0.85 | -2.96 | 0.40 | -1.48 | -1.57 | -2.56 | -0.21 | -2.50 | -0.57 |
| Sweden | -0.45 | 0.94 | 0.43 | 0.28 | 0.48 | 1.21 | 0.35 | -1.01 | 0.56 | 0.59 | 0.09 | 0.88 | 1.14 | 0.54 | 1.09 |
| Singapore | 1.13 | 0.88 | 1.37 | 1.37 | 1.27 | 0.98 | 0.68 | -0.50 | 1.14 | 0.37 | 1.08 | 1.45 | 0.65 | 0.06 | 0.10 |
| Slovenia | -0.25 | 0.59 | 0.11 | -0.42 | 0.97 | -0.54 | 0.28 | 0.58 | 1.90 | 0.18 | 0.27 | 0.36 | 0.99 | 0.52 | 1.04 |
| Slovakia | -0.55 | -0.98 | -0.03 | 0.33 | -0.29 | -0.65 | 0.79 | -0.80 | 0.25 | 1.11 | -0.30 | 0.80 | 0.64 | -0.36 | -0.05 |
| Turkey | 1.27 | -1.63 | -0.32 | -0.59 | 1.04 | -0.69 | 0.82 | 0.48 | 0.61 | 1.38 | 0.74 | -0.05 | -0.41 | 0.66 | 0.56 |
| United States | -0.17 | 0.42 | 0.01 | 0.64 | 0.31 | -0.76 | -0.28 | -0.21 | -0.23 | -0.15 | -0.04 | 0.06 | 0.05 | -0.40 | 0.18 |
| South Africa | -1.26 | 0.48 | -0.68 | -0.53 | -1.06 | 0.83 | -0.93 | -1.90 | -0.47 | -0.79 | -0.91 | -1.07 | -0.42 | -0.89 | -0.65 |